\documentclass[12pt]{article} 
\usepackage[fleqn]{amsmath}
\usepackage{ol2}
\usepackage[draft]{hyperref}
\usepackage{amsmath}

\begin{document}


\title{Cylindrical cloaking at oblique incidence with optimized finite multilayer parameters}


\author{Baile Zhang$^*$, and Bae-Ian Wu}

\address{
Research Laboratory of Electronics, Massachusetts
Institute of Technology, Cambridge, MA 02139, USA \\
$^*$Corresponding author: bzhang@mit.edu}

\begin{abstract}\ We propose multilayer cylindrical invisibility cloaks that are optimized for
oblique incidences through combination of analytic formalism of
scattering and genetic optimization. We show that by using only four
layers of homogeneous and anisotropic metamaterials without large
values of constitutive parameters, the scattering for oblique
incidences can be reduced by two orders. Although the optimization
is done at a single incident angle, the cloak provides reduced
scattering over a large range of incident angles.
\end{abstract}



\noindent

A lot of efforts have been made recently on designing and
implementing metamaterial invisibility cloaks using transformation
methods~\cite{pendry,leonhardt,cummer,hongsheng,baileprb,zhichao,schurig,wenshan,ying}.
A perfect invisibility cloak created from transforming empty
electromagnetic space~\cite{pendry} has been proven theoretically to
be perfect in cloaking an arbitrary object~\cite{hongsheng,
baileprb, zhichao}, while the only electromagnetic mechanism so far
to detect a perfect cloak within its working band needs extreme
conditions~\cite{baile_detection}. However, the practical
development of metamaterial invisibility cloaking is still far from
real application at this stage, since the rigorous requirements of
an ideal cloak, such as continuously varying parameters which
approach extreme values in some regions, are challenging for
practical fabrication. Although a simplified cylindrical cloak that
was implemented previously~\cite{schurig} can moderately reduce the
scattering, it is still inherently visible~\cite{minqiu_inherent},
requiring further improvements on its performance.

Moreover, previous studies on the cylindrical invisibility cloak,
which is often preferred in most analysis and experiments because of
its relatively simple geometry, were mostly limited to an in-plane
or purely two-dimensional (2D) case where the wave must propagate in
a 2D plane perpendicular to the cylindrical axis, which has severely
hindered the practicability of cylindrical cloaking. On the other
hand, recently, there have been theoretical suggestions to overcome
the drawbacks of continuous and extreme parameters by optimizing the
structure of a few layers of homogeneous and anisotropic
metamaterials to construct a practical
cloak~\cite{xi_optimization,popa_optimization}. However, similar to
previous studies, only normal incidence is considered and therefore
the results would only work for a purely 2D geometry. The extension
of a 2D design to a 3D design is one step further towards the
practically usable realization of invisibility cloaking.

We present an invisibility cloak design with only 4 layers of
metamaterials that can work at a wide range of incident angles. Our
work can be treated as the extension and supplement of the design of
purely 2D models into more physical and practical 3D models. We use
genetic optimization to globally search an optimized set of
parameters. The optimization of a multilayer cloak is done with
oblique incidence of 30 degrees. Our method can be used as a
guidance for future implementation of metamaterial invisibility
cloaks.

\begin{figure}
\centering
\includegraphics[width=0.5\columnwidth,draft=false]{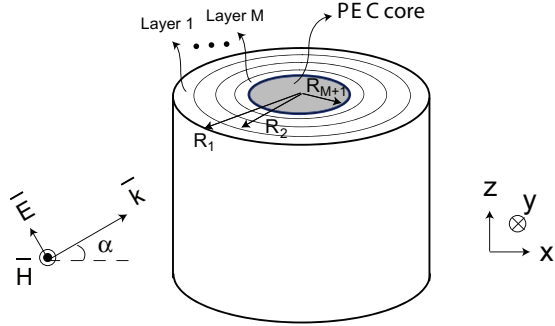}
\caption{\label{fig:geneticoblique} Configuration of a 4-layer cloak
when a vertically polarized plane wave is obliquely incident with
the incident angle of $\alpha$.}
\end{figure}

The geometry of the multilayer cylindrical cloak model is shown in
Fig.~\ref{fig:geneticoblique}, where a cylinder of perfect electric
conductor (PEC) with radius $R_{M+1}$ is cloaked by the multilayered
cylindrical shell with innermost radius $R_{M+1}$ and outermost
radius $R_1$. In between $R_1$ and $R_{M+1}$ are $M$ layers marked
by $m$ ($1\leq m \leq M$) with different outer radius varying from
$R_1$ to $R_M$. The constitutive parameters ($\epsilon_\rho$,
$\epsilon_\phi$, $\epsilon_z$, $\mu_\rho$, $\mu_\phi$ and $\mu_z$)
of each layer are assumed to be homogeneous. The region of $r>R_1$
is free space. The incident electromagnetic wave is incident on the
multilayer cylindrical cloak with an angle of $\alpha$ with respect
to the $xy$ plane. Without loss of generality, we consider vertical
polarization as shown in Fig.~\ref{fig:geneticoblique}. Therefore,
the unit incident wave takes the form of $\overline{E}_i =(-\hat x
\sin \alpha + \hat z \cos \alpha) e^{ikz \sin \alpha + ikx\cos
\alpha}$.

For oblique incidence, the difficulty here is that there is no
closed-form solution to the wave equation inside each layer. We have
provided a method to calculate the scattering from a general
cylindrical cloak under oblique incidence in \cite{baile_oe1} based
on the state-variable approach~\cite{weng}. We apply this method in
each layer to calculate each corresponding state propagator matrix,
respectively, and their product will be the final state propagator
matrix of the entire multilayer structure. If we divide each layer
into $N$ sublayers ($N\gg 1$), the state propagator equation will be
\begin{eqnarray}
&\overline V (R_{M+1}) = \nonumber & \\
&\left[ \prod_{j=(M-1)N+1}^{MN}(\overline{\overline I} + \Delta \rho
\overline{\overline T}(\rho_{j})) \right] & \nonumber
\\
&\cdot \left[ \prod_{j=(M-2)N+1}^{(M-1)N}(\overline{\overline I} +
\Delta \rho \overline{\overline T}(\rho_{j})) \right]&\nonumber
\\
&\cdot \cdot \cdot & \nonumber
\\
 & \cdot \left[
\prod_{j=1}^{N}(\overline{\overline I} + \Delta \rho
\overline{\overline T}(\rho_{j})) \right]&\nonumber \\
&\cdot \overline V (R_1)&
\end{eqnarray}
where the four-dimensional vector $\overline V = [E_z ~E_\phi ~H_z
~H_\phi]^T$ is the state vector. After obtaining the state
propagator matrix, we can solve the field distribution over the
entire space~\cite{baile_oe1}. Now by applying the genetic
optimization~\cite{xi_optimization}, a multilayer cloak applied at
oblique incidence is obtained.

For comparison with the performance of simplified cloaks at oblique
incidence, a cylindrical structure with the same size as in
\cite{baile_oe1} is considered, i.e. $R_1 = 1.5 \lambda_0 = 2.08
R_{M+1}$. We set the number of layers $M$ to be 4 where each layer
has the same thickness. Another condition is that we force all
relative constitutive parameters in each layer to be less than 10.
The incident angle of the incident wave is chosen to be $\alpha =
30^{\circ}$. In the genetic algorithm, the far-field scattering
efficiency $Q_{sca}$~\cite{baileprb}, or the scattering cross
section normalized by the unit geometrical cross section of
$2R_{M+1}$, is chosen as the function to be minimized.

\begin{figure}
\centering
\includegraphics[width=0.5\columnwidth,draft=false]{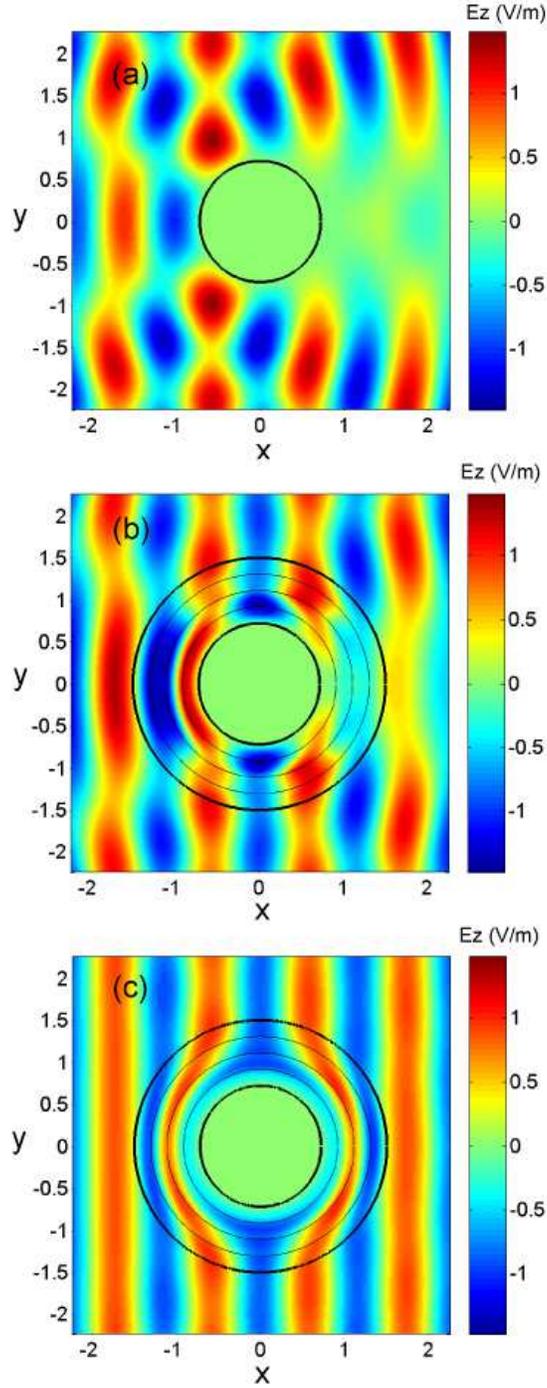}
\caption{\label{fig:total} (Color online) $E_z$ field distribution
in $xy$ plane due to scattering from (a) a bare PEC cylinder, (b) a
4-layer cloak with three parameters ($\epsilon_z$, $\mu_\rho$ and
$\mu_\phi$) optimized, and (c) a 4-layer cloak with all six
parameters optimized, when a vertically polarized plane wave is
obliquely incident from left to right with the incident angle of
$30^{\circ}$. From (a) to (c), the scattering efficiency $Q_{sca}$
is 1.175, 0.498 and 0.013, respectively.}
\end{figure}

We first plot the scattering from the bare PEC core (radius: $0.721
\lambda_0$) as the reference, where the $E_z$ field in the $xy$
plane is shown in Fig.~\ref{fig:total}(a).
 It is shown that the scattering is quite large. The
scattering efficiency is 1.175 in this case. Note that although the
incident wave is strictly vertically polarized, the reflected wave
is generally composed of both vertical and horizontal polarizations,
where only vertical polarization has $E_z$ component. The next task
is to minimize $Q_{sca}$ by optimization.

\begin{figure}
\centering
\includegraphics[width=0.5\columnwidth,draft=false]{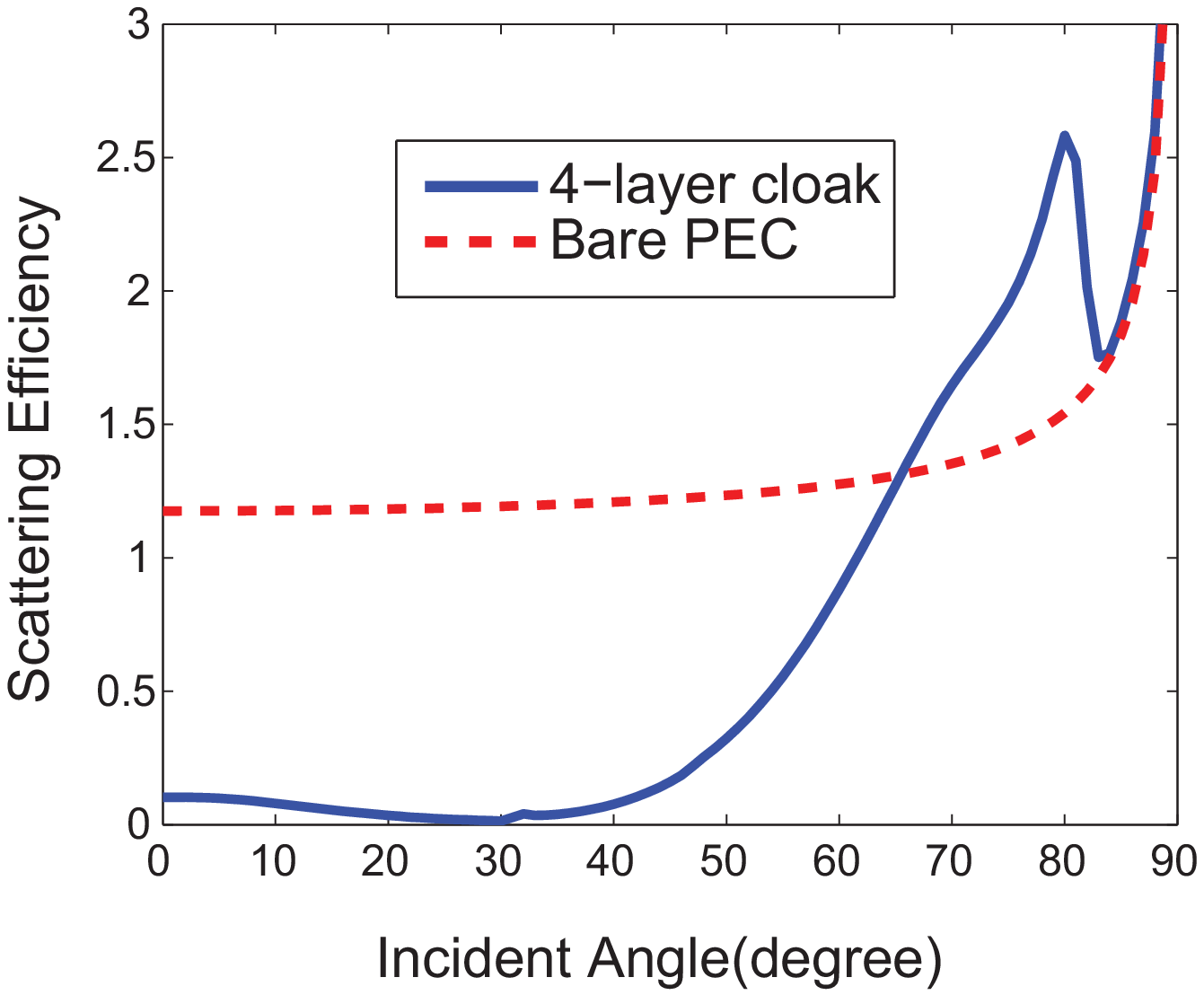}
\caption{\label{fig:RCSoblique} (Color online) Dependence of
scattering efficiency $Q_{sca}$ on the incident angle for the
4-layer cloak achieved by optimizing all six parameters in each
layer within the cloak. The dotted line corresponding to the
scattering efficiency of the bare PEC core is shown for comparison.}
\end{figure}

We have two methods of optimization. One is to control $\epsilon_z$,
$\mu_\rho$ and $\mu_\phi$ in each layer (the previous simplified
cloak for normal incidence only need these three parameters), while
keeping $\epsilon_\rho$, $\epsilon_\phi$ and $\mu_z$ as constant of
1. The other is to control all six constitutive parameters in each
layer. The performances of these two optimized 4-layer cloaks are
shown in Fig.~\ref{fig:total}(b-c). Their relative constitutive
parameters are summarized in Table~\ref{table:obliqueparameters}. It
can be seen in Fig.~\ref{fig:total}(b) that by controlling only
three parameters in each layer, the total scattering has reduced
much when compared to the bare PEC core in Fig.~\ref{fig:total}(a).
The scattering efficiency $Q_{sca}$ is reduced from 1.175 to 0.498.
After optimizing all six parameters in each layer in the cloak, the
scattering efficiency $Q_{sca}$ is 0.013, close to ``perfect
invisibility'' as shown in Fig.~\ref{fig:total}(c). It is worth
mentioning that although the 4-layer cloak in
Fig.~\ref{fig:total}(c) is achieved by optimization at the single
incident angle of $30^\circ$, it is still valid for other incident
angles. In Fig.~\ref{fig:RCSoblique}, we can see that this cloak has
satisfactory performance over a large range of incident angles.

\begin{table}
\begin{center}
\begin{tabular}{|c|ccc|ccc|}
\hline
 I&\multicolumn{6}{c|}{Optimizing three parameters} \\
\hline
 layer & $\mu_{\rho}$ & $\mu_{\phi}$ & $\epsilon_{z}$  & $\epsilon_{\rho}$ & $\epsilon_{\phi}$ & $\mu_{z}$ \\
\hline
1 & 0.962 & 1.423 & 1.154 & 1 & 1 & 1\\
 2 & 1.128 &0.160  &  0.598 & 1 & 1 & 1\\
 3 & 0.792 & 2.754 & 1.321 &1 &  1 & 1\\
 4 & 0.665 & 1.025 & 2.625 &1 &  1 & 1 \\
 \hline
  II&\multicolumn{6}{c|}{Optimizing six parameters} \\
\hline
 layer & $\mu_{\rho}$ & $\mu_{\phi}$ & $\epsilon_{z}$  & $\epsilon_{\rho}$ & $\epsilon_{\phi}$ & $\mu_{z}$ \\
\hline
1 & 0.550 & 2.320 & 1.725 & 0.438 & 1.781 & 3.409\\
 2 & 0.283 & 3.584 & 1.657 &  0.262 & 3.953 & 0.037\\
 3 & 0.224 & 8.204 & 0.614& 0.189 & 8.674 & 1.754\\
 4 & 0.057 & 9.994 & 0.015 & 1.540 & 9.069 & 0.030 \\
\hline
\end{tabular}
\caption{\label{table:obliqueparameters} The relative constitutive
parameters for the optimized 4-layer cloak by (I) optimizing three
parameters in each layer and (II) optimizing six parameters in each
layer.}
\end{center}
\end{table}

In principle, we can ascribe this low scattering phenomenon to the
destructive interference between the waves from the first reflection
on the outermost boundary and the waves from the total transmission
from inside the cloak to outside~\cite{xi_optimization}. However,
the scenario of oblique incidence is more complicated than the case
of normal incidence, because both polarizations need to be optimized
at the same time. For normal incidence, since vertically polarized
waves and horizontally polarized waves are decoupled, reflection and
transmission at individual boundaries between adjacent layers will
affect one polarization without influence from the other
polarization. Therefore it is relatively easier to achieve
destructive interference outside the cloak.  When the incident angle
is nonzero, vertically polarized and horizontally polarized waves
become coupled. Thus there is a need for more degrees of freedom to
optimize the structure to achieve destructive interference for both
polarizations. This justifies a much better performance of the
4-layer cloak with all constitutive parameters optimized in each
layer (Fig.~\ref{fig:total}(c)), as oppose to just 3 parameters
(Fig.~\ref{fig:total}(b)).

In conclusion, we have proposed a method to achieve multi-layered
cylindrical cloaks working at oblique incidence by combining the
analytic algorithm of scattering and genetic optimization. We show
that by using only four layers of homogeneous and anisotropic
metamaterials with constitutive parameters of finite values, the
scattering from the cloak at oblique incidence can be reduced by two
orders. Although the optimization is performed only at a single
incident angle, the performance of the cloak is satisfactory within
a large range of incident angles. We anticipate that the combination
of analytic algorithm of scattering and genetic optimization as a
powerful tool will be used in more applications in the practical
implementation of invisibility cloaks in the future.


\end{document}